\documentclass[aps,prl,groupedaddress,reprint,floatfix,notitlepage,nobalancelastpage,nopacs]{revtex4-1}
\usepackage{graphicx,amsmath,amssymb,bm,color,engord,nicefrac,braket}
\usepackage[usenames,dvipsnames]{xcolor}
\usepackage[colorlinks,colorlinks,urlcolor=Magenta,citecolor=Magenta,linkcolor=black]{hyperref}
\begin{document}

\title{A fully controllable Kondo system: Coupling a flux qubit and an ultracold Fermi gas}
\author{Kelly R. Patton}
\email[\hspace{-1.4mm}]{kpatton1@ggc.edu}
\affiliation{School of Science and Technology, Georgia Gwinnett College, Lawrenceville, GA 30043, USA }
\date{\today}
\begin{abstract}
We show that  a composite spin-\nicefrac{1}{2} Kondo system can be formed by coupling a superconducting quantum interference device (SQUID) to the internal  hyperfine states of a trapped ultracold atomic Fermi gas. Here, the SQUID, or flux qubit, acts as an effective magnetic impurity that induces  spin-flip scattering near the Fermi energies of the trapped gas. Although the ultracold gas and SQUID  are at vastly different temperatures,  the formation of a strongly correlated Kondo state between the two systems is found when the gas is cooled below the Kondo temperature;  this temperature regime is  within current experimental limits.  Furthermore,  the  momentum distribution of the trapped fermions is calculated.  We find that it clearly contains an experimental signature of this correlated state and the associated Kondo screening length.  In addition to probing Kondo physics, the controllability of this system can be used to systematically explore the relaxation and equilibration of a strongly correlated system that has been initially prepared in a selected  nonequilibrium state.
\end{abstract}
\maketitle

In the mid-1960s Jun Kondo gave the first detailed theoretical explanation of the anomalously large low-temperature resistivity that was observed in some metals \cite{KondoPTP64}. This effect,  which now bears his name, occurs when conduction electrons scatter off localized magnetic impurities with internal spin degrees of freedom.  He showed that this  exchange interaction  leads to a breakdown of perturbation theory below an energy scale: the so-called Kondo temperature.  This phenomenon was later understood in detail within Wilson's renormalization group \cite{WilsonRMP75,AndersonJPhysC70}.  In the renormalization sense, below the Kondo temperature an initial arbitrarily weak exchange coupling between the impurity and conduction electrons flows to the strong coupling regime. In the antiferromagnetic case, this flow ultimately produces a quasi-bound impurity state at the Fermi energy. It is the formation of this bound state that gives raise to the breakdown of perturbation theory and the enhancement of resistive scattering \cite{NozieresJLTP74}.  

The advent of  heterogeneous semiconductor devices brought about a dramatic resurgence of interest in the Kondo effect in the late-90s \cite{GoldhaberNatrue98,CronenwettScience98}.  In such systems, artificial magnetic impurities atoms are created, manipulated, and coupled to Fermi gas  leads.  This enabled the controlled  study of single and multiple Kondo impurities \cite{JeongScience01} in parameter regimes previously unexplored, including:  magnetic field dependence,  nonequilibrium effects, and non-Fermi liquid ground states  \cite{Kellerarxiv15}.  More recently, scanning tunneling microscopes (STM) \cite{MadhavanScience98,KnorrPRL02} and spin-polarized STMs \cite{vonBergmannPRL15} have imaged the  Kondo effect produced by single magnetic adatoms, chains \cite{DiLulloNanoLett12}, and  corrals \cite{ManoharanNature00}.  While not as intrinsically tunable as quantum dots, coupling the impurity to an exotic bath state, such as a spin-imbalanced or superconductive state, is more easily realizable in adatoms systems \cite{TernesJPhysCondMat09}. 
\begin{figure}
\includegraphics[scale=0.4]{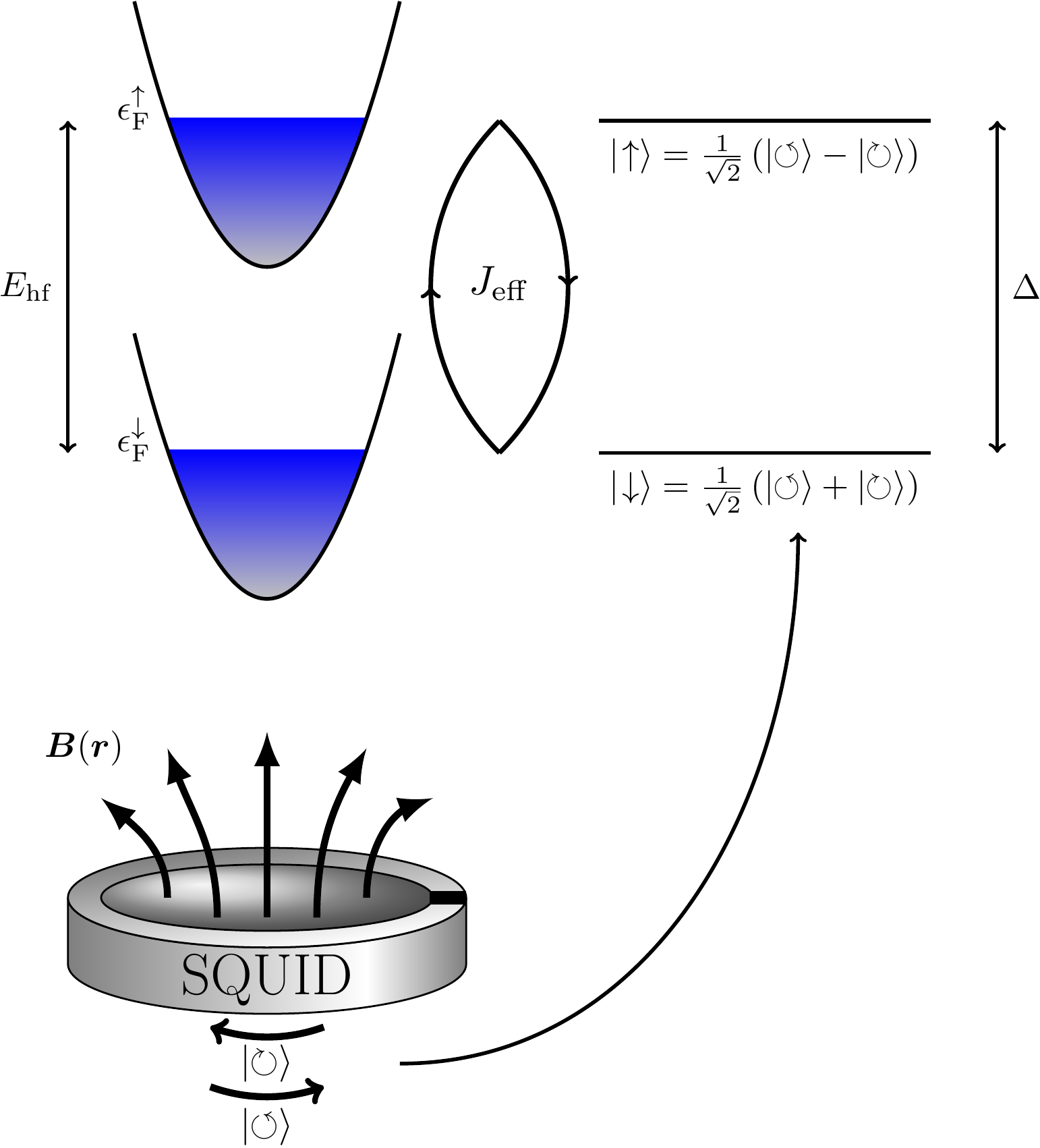}%
\caption{Schematic of the proposed composite Kondo system.  The  low-energy Hamiltonian of the SQUID can be modeled by an effective spin-\nicefrac{1}{2} system with Zeeman splitting $\Delta$. The two spin states of the Fermi gas correspond to  different  total atomic angular momentum states, with a hyperfine energy splitting $E_{\rm hf}$.  The magnetic field ${\bm B}({\bm r})$ produced by the clockwise $\ket{\circlearrowright}$ and counter-clockwise $\ket{\circlearrowleft}$ supercurrent states of the SQUID induces an  exchange interaction $J_{\rm eff}$ between the flux qubit and Fermi gas. When on resonance, $\Delta\approx E_{\rm hf}$,  this interaction efficiently produces spin flips between the impurity and  bath states via the Rabi cycle.  \label{fig1}}
\end{figure}

Here, we propose a hybrid  Kondo system whose every aspect is essentially  fully controllable and tunable. The  dimensionally, lattice structure, spin imbalance,  interactions, and  superconducting state of the Fermi gas can all be externally manipulated.  The spin impurity itself is also tunable.  Its effective Zeeman splitting can be dynamically controlled, as well as its  position on the Bloch sphere.  Furthermore, the gas-impurity exchange interaction can be tuned from weak to strong coupling.  The proposed system is formed  by coupling a superconducting quantum interference device (SQUID), or flux qubit, to a two-component ultracold atomic Fermi gas.  The low-energy  Hilbert space of the SQUID is described by an effective Zeeman split  spin-\nicefrac{1}{2} Hamiltonian.  The two effective spin states are actually symmetric and antisymmetric combinations of left and right circulating macroscopic superconducting currents. These currents  produce a magnetic field and, when on resonance,  can induce magnetic dipole transitions between two selected hyperfine spin states of the trapped gas, see Fig.~\ref{fig1}. Below we show that the combined system can be described by a Kondo-like Hamiltonian.  Similar setups involving an atomic Bose-Einstein condensate have been recently put forth and realized experimentally \cite{PattonPRA13,PattonPRL13,WeissPRL15}.

A SQUID is  a superconducting ring interrupted by a thin tunneling barrier. An applied magnetic flux induces clockwise $\ket{\circlearrowright}$ or counterclockwise $\ket{\circlearrowleft}$ screening supercurrents in response.  The current states are not energy eigenstates, but instead are eigenstates of the current operator $\hat{I}=I\sigma_{z}$ with eigenvalues $\pm I$,  corresponding to the total  current and direction of each state.  The two lowest energy eigenstates of the SQUID are $\ket{\downarrow}_{\rm S}=2^{-1/2}(\ket{\circlearrowleft}+\ket{\circlearrowright})$ and $\ket{\uparrow}_{\rm S}=2^{-1/2}(\ket{\circlearrowleft}-\ket{\circlearrowright})$.  In this basis the Hamiltonian of the flux qubit can be written as $(\hbar =1)$ $H_{\rm S}=\frac{\Delta}{2}\sigma_{z}$; typically $\Delta\sim 1$--$10\,{\rm GHz}$ \footnote{See Ref. \cite{PattonPRA13} for details.}.   Here the energy level spacing $\Delta$ will remain fixed, but in practice  can be dynamically controlled on sub-nanosecond times scales \cite{PaauwPRL09,CastellanoNJP10}.   The macroscopic magnetic field ${\bm B}({\bm r})$ generated by the circulating currents can be determined  from the Biot-Savart law, given the SQUID's geometry and current $I$. In the current basis the magnetic field operator  is  ${\hat{\bm B}}({\bm r})={\bm B}({\bm r})\sigma_{z}$.

The fermion-fermion interaction of the two-component trapped gas can be externally tuned (through a Feshbach resonance) from strongly to weakly interacting. For simplicity we'll assume they are in the weakly interacting regime and above any superfluid transition temperature \cite{BlochRMP08}.  
The two experimentally prepared hyperfine states are atomic states of total internal angular momentum ${\bm F}={\bm J}+{\bm I}$ with $z$-projection  $m_{F}$ and corresponding internal energies $\omega_{\sigma}$ that we label with a pseudo-spin-\nicefrac{1}{2} index $\sigma=\uparrow,\downarrow$.  For example, in $^{40}{\rm K}$ the two states could be $\ket{\uparrow}=\ket{F=7/2, m_{F}=-7/2}$ and $\ket{\downarrow}=\ket{F=9/2, m_{F}=-7/2}$, which have a hyperfine splitting of $E_{\rm hfs}=\omega_{\uparrow}-\omega_{\downarrow}\approx 1.3\,{\rm GHz}$.  

An interaction between the SQUID and the atomic gas is induced by the magnetic field produced by the flux qubit.  This coupling can induce magnetic-dipole driven spin flips between the two hyperfine spin states and the two  spin states of the SQUID.  Neglecting coupling to other internal states, this interaction is described by $V=-\sum_{\sigma,\sigma'}\int {\rm d}{\bm r}\,\hat{\Psi}^{\dagger}_{\sigma}({\bm r})\boldsymbol{\mu}_{\sigma,\sigma'}\hat{\Psi}_{\sigma'}({\bm r})\otimes \hat{\bm B}({\bm r})$,
where $\boldsymbol{\mu}^{}$ is the Land\'{e} $g$-factor weighted sum of each internal angular momentum contribution to the total magnetic moment of the atom. In the weak-field limit  $\boldsymbol{\mu}^{}\approx g^{}_{F}\mu^{}_{\rm B}\bm F$, where $\mu^{}_{\rm B}$ is the Bohr magneton.   Decomposing the electron field operators in terms of the mode operators of the trapped gas $\hat{\Psi}^{(\dagger)}_{\sigma}({\bm r})=\sum_{{\bm n}}\phi^{(*)}_{{\bm n}}({\bm r})\hat{c}^{(\dagger)}_{{\bm n}\sigma}$ and changing to the energy eigenstate basis of the isolated flux qubit ($\ket{\uparrow}$ and $\ket{\downarrow})$, the total  Hamiltonian can then be expressed as
\begin{equation}
\label{Exact Hamiltonian}
H=\sum_{{\bm n},\sigma}\xi^{}_{{\bm n}\sigma}\hat{c}^{\dagger}_{{\bm n}\sigma}\hat{c}^{}_{{\bm n}\sigma}+\frac{\Delta}{2}\sigma^{}_{z}+\sum_{{\bm n},{\bm n}'}\sum_{\sigma,\sigma'}J^{}_{{\bm n}\sigma,{\bm n}'\sigma'}\hat{c}^{\dagger}_{{\bm n}\sigma}\hat{c}^{}_{{\bm n}'\sigma'}\sigma_{x},
\end{equation}
where $\xi^{}_{{\bm n}\sigma}=\epsilon_{\bm n}+\omega_{\sigma}$ are the single-particle fermion energies and $J^{}_{{\bm n}\sigma,{\bm n}'\sigma'}=\sum_{i=x,y,z}J^{i}_{{\bm n},{\bm n}'}F^{i}_{\sigma,\sigma'}$
with $J^{i}_{{\bm n},{\bm n}'}=-g^{}_{F}\mu^{}_{\rm B}\int {\rm d}{\bm r}\,\phi^{*}_{\bm n}({\bm r})B^{i}({\bm r})\phi^{}_{{\bm n}'}({\bm r})$ is the effective exchange coupling.  Equation \eqref{Exact Hamiltonian} is the Hamiltonian  of a Kondo-like system with a highly anisotropic exchange interaction and  an applied Zeeman field.  Typically, a Zeeman field tends to  suppress the Kondo effect.  Here, the effective Zeeman splitting $\Delta$ of the impurity states is essential to its formation in this system, as the spin flip scattering is primarily driven by the Rabi process.  During a Rabi cycle, the effects of the interaction are maximized when the two systems are on resonance, i.e., when $E_{\rm hfs}= \Delta$, and highly suppressed when far-off resonance. 
Although the Hamiltonian, Eq.~\eqref{Exact Hamiltonian}, describes a Kondo-like system, it is far from  thermal equilibrium,   as the SQUID's typical operating temperature  is on the order of a millikelvin, while the Fermi gas  can be in the nanokelvin regime.  As we will show in the following, the Kondo regime, or temperature, is  set by the temperature of the gas. 

The nonequilibrium  contour-ordered bath Green's function is
$ G^{c}_{{\bm n}\sigma,{\bm n}'\sigma'}(z,z')=-i\langle {\cal T}^{}_{C}\hat{c}^{}_{{\bm n}\sigma}(z)\hat{c}^{\dagger}_{{\bm n}'\sigma'}(z')\rangle$,
where $C=C_{+}\cup C_{-}$ is the Keldysh time contour with  upper $C_{+}$ and lower $C_{-}$ branches \cite{NEGFbook}. The expectation value  is with respect to the non-interacting density matrix, while the time-evolution of the operators is with respect to the full Hamiltonian \footnote{Technically we have to work in the canonical ensemble as we require equal numbers of spin up and down  atoms even though their energies are offset by the hyperfine splitting.}.  The contour Green's function can also be expressed in matrix form as
\begin{equation*}
{\bf G}_{{\bm n}\sigma,{\bm n}'\sigma'}(t,t')=\left(\begin{array}{cc}G_{{\bm n}\sigma,{\bm n}'\sigma'}(t,t') & G^{<}_{{\bm n}\sigma,{\bm n}'\sigma'}(t,t') \\G^{>}_{{\bm n}\sigma,{\bm n}'\sigma'}(t,t') & \tilde{G}_{{\bm n}\sigma,{\bm n}'\sigma'}(t,t')\end{array}\right),
\end{equation*}
where $G(t,t)$ is the standard time-ordered Green's function, $\tilde{G}(t,t)$ the anti-time-ordered one, and $G^{\lessgtr}(t,t')$ are the lesser and greater correlation functions.  Furthermore, to enable standard diagrammatic techniques  we fermionize the impurity spin operators \cite{AbrikosovPhysica65,ZawadowskiZPhysik69} \footnote{This process enlarges the Hilbert space to include an unphysical doubly occupied impurity state. Thus,  any impurity correlation function must be eventually projected back into the physically meaningful Hilbert space.}.

\begin{figure}
\includegraphics[scale=0.6]{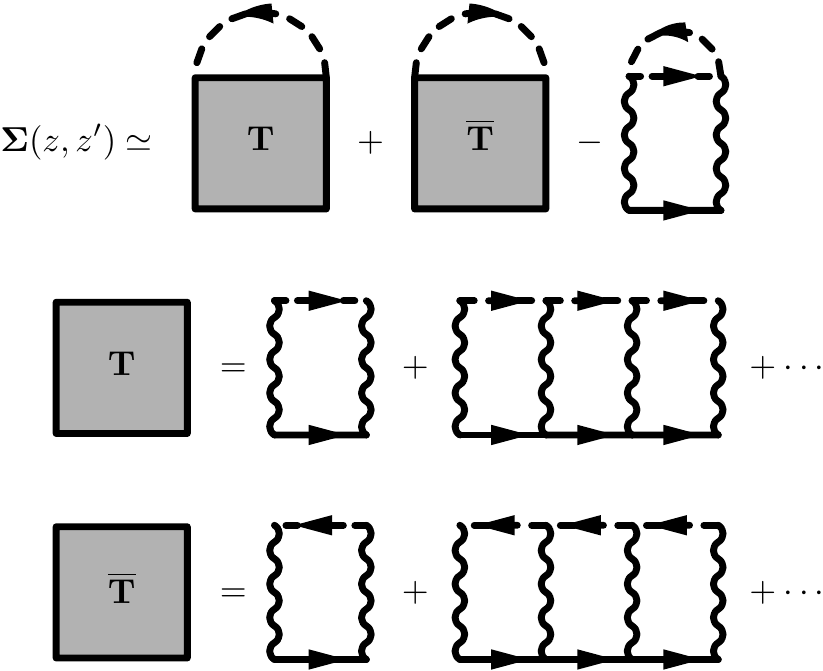}%
\caption{The diagrams included in the particle-particle and particle-hole symmetric ladder approximation, or $T$-matrix, to the self-energy of the Fermi gas. The solid line represent the fermion propagator of the gas,  while the dashed line is the projected pseudo-fermion propagator of the impurity. The spin and state dependent interaction between the two is denoted by the wavy line.  The subtraction of the last diagram in the self-energy is to avoid over counting the leading order term, which can be expressed either channel.  \label{fig2}}
\end{figure}

Although the  Kondo effect ultimately leads to a breakdown of perturbation theory, perturbative results  remain quantitatively correct down to energy  scales on the order of the Kondo Temperature $T_{\rm K}$.  To this end, we sum a subset of the leading-order logarithmically divergent diagrams,  maintaining particle-hole symmetry, for the fermion self-energy \cite{LebanonPRB01}.  The approximate self-energy is found in the so-called $T$-matrix approximation, or ladder series,  in both the particle-particle and particle-hole channels, see Fig.~\ref{fig2}. 
The rung of each ladder series describes the scattering of a bath  and impurity fermion or hole.  Within the steady-state approximation and Fourier transformed, the Keldysh matrix components of the rung of the $T$-matrix in the  particle-particle channel is  $\Pi^{ij}_{{{\bm n}\sigma s;{\bm n}'\sigma's'}}(\omega)=\int \frac{{\rm d}\omega'}{2\pi}J^{s,s'}_{{\bm n}\sigma,{\bm n}'\sigma'} G^{ij}_{{\bm n'}\sigma'}(\omega-\omega')G^{ij}_{s'}(\omega')$,
 where $J^{s,s'}_{{\bm n}\sigma,{\bm n}'\sigma}=J^{}_{{\bm n}\sigma,{\bm n}'\sigma}\sigma^{s,s'}_{x}$ and in the particle-hole channel $\overline{\Pi}^{ij}_{{{\bm n}\sigma s;{\bm n}'\sigma's'}}(\omega)=\int \frac{{\rm d}\omega'}{2\pi}J^{s,s'}_{{\bm n}\sigma,{\bm n}'\sigma'} G^{ij}_{{\bm n'}\sigma'}(\omega+\omega')G^{ji}_{s'}(\omega'),$ where $G^{}_{{\bm n}\sigma}$ is the non-interacting  Green's function of the bath and $G^{}_{s}$ is the {\it projected} non-interacting  pseudo-fermion Green's function.

The fermion self-energy in this approximation is explicitly given by
\begin{align*}
\Sigma^{ij}_{{\bm n}\sigma,{\bm n}'\sigma'}(\omega)&=\sum_{s}\int \frac{{\rm d}\omega'}{2\pi}T^{ij}_{{{\bm n}\sigma s;{\bm n}'\sigma's}}(\omega+\omega')G^{ji}_{s}(\omega')\nonumber\\&+\sum_{s}\int \frac{{\rm d}\omega'}{2\pi}\overline{T}^{ij}_{{\bm n}\sigma s;{\bm n}'\sigma's}(\omega-\omega')G^{ij}_{s}(\omega')\nonumber\\&-\sum_{s}\sum_{{\bm n}_{1}\sigma_{1}s_{1}}\int \frac{{\rm d}\omega'}{2\pi}\Pi^{ij}_{{{\bm n}\sigma s;{\bm n}_{1}\sigma_{1}s_{1}}}(\omega+\omega')\nonumber\\&\times J^{s_{1},s}_{{\bm n}_{1}\sigma_{1},{\bm n}'\sigma'}G^{ji}_{s}(\omega'),
\label{self energy}
\end{align*}
where the $T$-matrices satisfy the following integral equations in Keldsyh space
\begin{widetext}
\begin{equation*}
\label{pp-T-matrix}
{\bf T}_{{\bm n}\sigma s;{\bm n}'\sigma's'}(\omega)=\sum_{{\bm n}_{1}\sigma_{1}s_{1}}\,\boldsymbol{\Pi}_{{\bm n}\sigma s;{\bm n}_{1}\sigma_{1}s_{1}}(\omega)J^{s_{1},s'}_{{\bm n}_{1}\sigma_{1},{\bm n}'\sigma'}-\sum_{\substack{{\bm n}_{1},{\bm n}_{2}\\ \sigma_{1},\sigma_{2}}}\sum_{s_{1},s_{2}}\boldsymbol{\Pi}_{{\bm n}\sigma s;{\bm n}_{1}\sigma_{1}s_{1}}(\omega)\sigma_{z}\boldsymbol{\Pi}_{{\bm n}_{1}\sigma_{1} s_{1};{\bm n}_{2}\sigma_{2}s_{2}}(\omega)\sigma_{z}{\bf T}_{{\bm n}_{2}\sigma_{2} s_{2};{\bm n}'\sigma's'}(\omega)
\end{equation*}
and
\begin{equation*}
\label{ph-T-matrix}
{\bf \overline{T}}_{{\bm n}\sigma s;{\bm n}'\sigma's'}(\omega)=\sum_{{\bm n}_{1}\sigma_{1}s_{1}}\,\boldsymbol{\overline{\Pi}}_{{\bm n}\sigma s;{\bm n}_{1}\sigma_{1}s_{1}}(\omega)J^{s_{1},s'}_{{\bm n}_{1}\sigma_{1},{\bm n}'\sigma'}-\sum_{\substack{{\bm n}_{1},{\bm n}_{2}\\ \sigma_{1},\sigma_{2}}}\sum_{s_{1},s_{2}}\boldsymbol{\overline{\Pi}}_{{\bm n}\sigma s;{\bm n}_{1}\sigma_{1}s_{1}}(\omega)\sigma_{z}\boldsymbol{\overline{\Pi}}_{{\bm n}_{1}\sigma_{1} s_{1};{\bm n}_{2}\sigma_{2}s_{2}}(\omega)\sigma_{z}{\bf \overline{T}}_{{\bm n}_{2}\sigma_{2} s_{2};{\bm n}'\sigma's'}(\omega).
\end{equation*}
\end{widetext}

Within this approximation the  the  Kondo temperature will be taken as the temperature at which the $T$-matrices diverge; this signals the breakdown of perturbation theory.  To make further progress  we neglect the state dependence of the exchange interaction; $J^{s,s'}_{{\bm n}\sigma,{\bm n}'\sigma'}\to J^{s,s'}_{\sigma,\sigma'}\approx J/{\sf V}(\sigma^{\sigma,\sigma'}_{x}+\sigma^{\sigma,\sigma'}_{y})\sigma^{s,s'}_{x}$, where $J$ is  taken to be the value of the coupling for states near the Fermi surface and  ${\sf V}$ is the volume of the trapped gas.  Additionally, for now we set the SQUID's temperature to be much larger than it's level spacing. This essentially makes the two impurity states degenerate.  One can then show that the above perturbation series diverges when the temperature of the gas $k_{\rm B}T_{\rm g}$ approaches  $k_{\rm B}T_{\rm K}\sim \epsilon_{\rm F}\exp\left[-1/(2|J|\rho_{0})\right]$, where $\rho_{0}$ is the single-particle density of states per spin at the Fermi energy $\epsilon_{\rm F}$.  Estimating the exchange coupling $J$ for a two-dimensional gas with with Fermi wavelength $\lambda_{\rm F}\approx 1\,\mu{\rm m}$ and  Fermi energy $\epsilon_{\rm F}\approx 1\,\mu{\rm K}$ that is placed $100\,\mu {\rm m}$ above a $1\,\mu{\rm m}$ radius SQUID carrying $1\,{\rm mA}$ of current gives $k_{\rm B}T_{\rm K}\sim 0.01 \epsilon_{\rm F}$, which is within current experimental limits. Although the inclusion of the momentum dependence can change the functional relationship between the coupling and Kondo energy scale, this estimate should still be qualitatively correct for weak coupling.  One remaining uncertainty is the effect of  decoherence  on the formation of the Kondo state \cite{KatsnelsonPLA09}. A detailed treatment of this will be left for future investigation,  a simple estimate can be made by noting that  the energy relaxation time of a current flux qubit $T_{1}\sim1\,\mu {\rm s}$ and the time required for the Kondo singlet to form  $t_{\rm K}\sim (k_{\rm B}T_{\rm K})^{-1}$ \cite{NordlanderPRL99} are of the same order.  Thus, it should be possible for correlations to form before the SQUID decoheres. 

\begin{figure}
\includegraphics[scale=0.8]{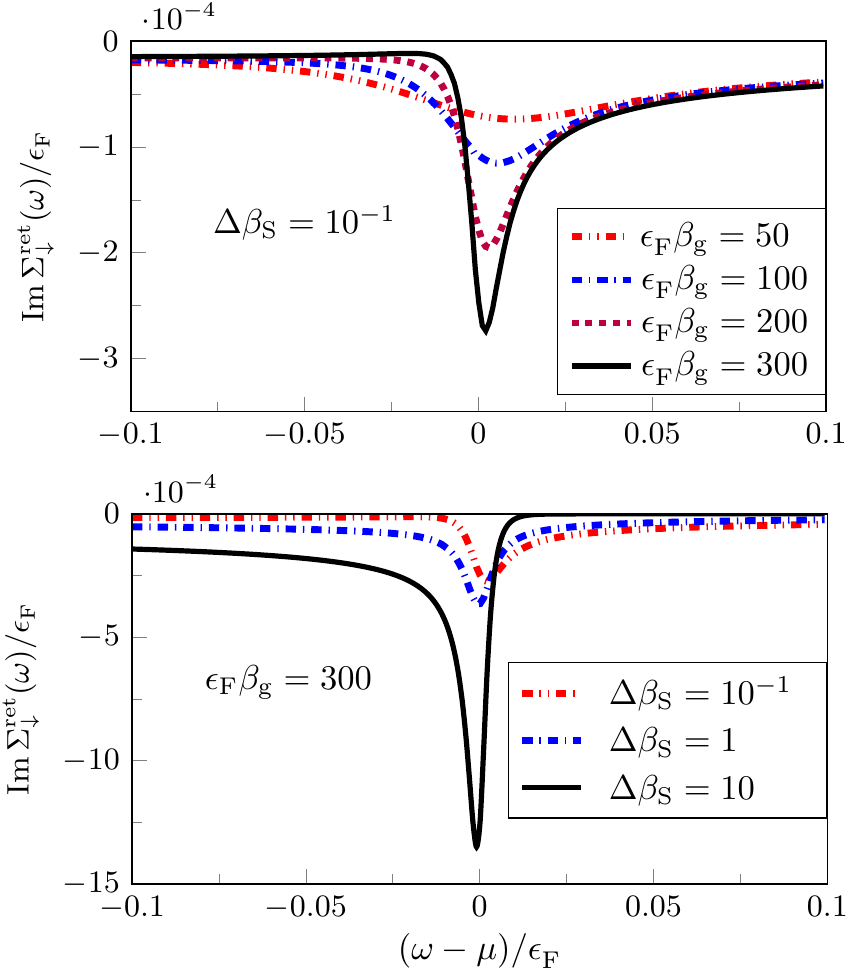}%
\caption{The imaginary part of the retarded self-energy, which is proportional to the scattering rate,  for the spin-$\downarrow$ atomic fermions  is shown for various temperature regimes with $\beta_{\rm S}$ and $\beta_{\rm g}$ being the inverse temperatures of the SQUID and atomic gas respectively.  The top panel shows that as the temperature of the gas crosses the Kondo temperature a Kondo resonance forms near the Fermi energy $\epsilon_{\rm F}=\mu$. The Kondo temperature for the system is $\epsilon_{\rm F}\beta_{\rm K}\sim 150$. The bottom panel shows how the scattering rate depends on the temperature of the SQUID when the Fermi gas is well below the Kondo temperature. When $\Delta \beta_{\rm S}\ll 1$ the spin states of the SQUID are essentially degenerate. As the temperature of the SQUID is lowered this gives raise to an effective magnetization of the impurity.  \label{fig3}}
\end{figure}

\begin{figure}
\includegraphics[width=\columnwidth]{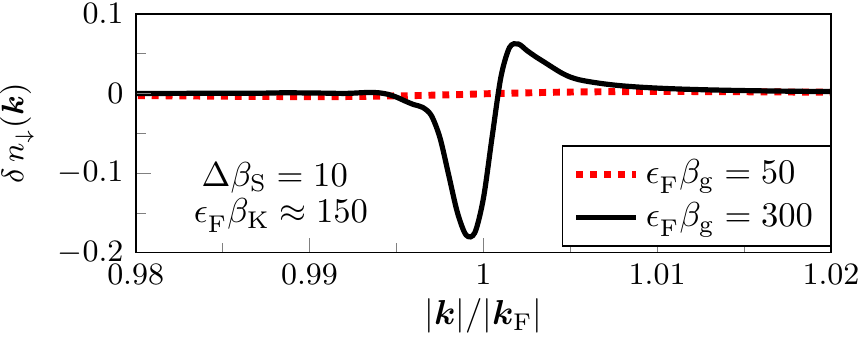}%
\caption{The change in the momentum distribution of the atomic gas induced by Kondo correlations for a system far above the Kondo temperature (dashed line) and well below (solid line).  In real space these correlations have a spatial extent given by the so-called Kondo screening length $\xi^{}_{\rm K}=v_{\rm F}\beta_{\rm K}$, where $v_{\rm F}$ is the Fermi velocity and $\beta_{\rm K}=(k_{\rm B}T_{\rm K})^{-1}$. The associated momentum scale $|{\bm k}_{\rm K}|=2\pi/\xi_{\rm K}$ defines the thin shell of momentum states near the Fermi surface which take part in the screening, i.e., those states with ${\bm k}={\bm k}_{\rm F}\pm {\bm k}_{\rm K}$.   \label{fig4}}
\end{figure}

The Kondo effect can manifest itself as a large increase in the scattering rate, or inverse lifetime, of the bath fermions near the Fermi energy as the temperature approaches the Kondo regime. The scattering rate  is proportional to the imaginary part of the retarded self-energy $\Sigma^{\rm ret}$, which can be obtained from the lessor and greater self-energies.   Figure \ref{fig3} shows the imaginary part of the retarded self-energy for the spin-$\downarrow$ atomic fermions in various temperature regimes of the gas and the SQUID \footnote{By neglecting the state dependence in the exchange coupling, the fermion self-energy becomes inversely proportional to the volume of the trapped gas.  In the thermodynamic limit this volume factor would be replaced by a finite impurity concentration.    Here, the volume of the gas is finite and is taken to be a two-dimensional square with a side length of ten Fermi wavelengths.}.  As the temperature of the gas is lowered into the Kondo regime, the lifetime of the quasiparticles near the Fermi surface dramatically decreases.  In a typical condensed matter system this enhancement of the scattering rate would express itself as an increased resistance.    
Measuring such transport properties in ultracold atomic systems is very challenging, but the momentum distribution of the gas can be easily obtained by free expansion \cite{BlochRMP08}.  We find the momentum distribution also contains a signature of the Kondo effect and the so-called Kondo screening length \cite{AffleckPRL01}. Roughly, this length scale sets the spatial extent of the singlet state around the impurity.  To date,  this length has  never been experimentally observed.   The momentum distribution of the atomic gas is given in terms of the spectral function;  $n_{\sigma}({\bm k})=\int \frac{{\rm d}\omega}{2\pi }A_{\sigma,\sigma}({\bm k},{\bm k},\omega)n_{\rm F}(\omega-\mu_{\sigma}),$
where  $A_{\sigma,\sigma'}({\bm k},{\bm k},\omega)=-2\,{\rm Im}\, G^{\rm ret}_{{\bm k}\sigma,{\bm k}'\sigma'}(\omega)$ and $n_{\rm F}(\omega)$ is the Fermi distribution of the gas.  To leading order in the self-energy this can be written as
\begin{align}
\label{approximate momentum distribution}
&\delta n_{\sigma}({\bm k})=n_{\sigma}({\bm k})-n^{}_{\rm F}(\xi_{{\bm k}\sigma}-\mu_{\sigma})\nonumber\\&-\frac{1}{\pi}\frac{{\rm d}}{{\rm d}\xi_{{\bm k}\sigma}}\int {\rm d}\omega\,{\rm Im}[\Sigma^{\rm ret}_{\sigma,\sigma}(\omega)]n_{\rm F}(\omega-\mu_{\sigma}){\rm P}\left(\frac{1}{\omega-\xi_{{\bm k}\sigma}}\right).
\end{align}
Figure \ref{fig4} shows the momentum distribution above and below the Kondo temperature.  A qualitative  understanding of these results can be had as follows: Deep in the Kondo regime the imaginary part of the self-energy is strongly peaked near the Fermi energy. The width of this peak is of order the inverse Kondo temperature $\beta^{-1}_{\rm K}$. Thus, the integral in Eq.~\eqref{approximate momentum distribution} will only vary significantly when ${\bm k}\approx {\bm k}_{\rm F}\pm {\bm k}_{\rm K}$, where $|{\bm k}_{\rm K}|=2\pi/\xi_{\rm K}$ and $\xi_{\rm K}=v_{\rm F}\beta_{\rm K}$ is the Kondo screening length.  Therefore, only momentum states within this shell of the Fermi surface participate in forming the Kondo cloud.  

We have shown that a fully controllable hybrid Kondo system can be formed by coupling a flux qubit to an ultracold atomic Fermi gas. The high degree of tunability of both the bath and the impurity offers the prospect of exploring the Kondo effect in regimes never before realized.  Critical to this---we find that the Kondo temperature for this system is within current experimental limits. The detection of the strongly correlated state between the impurity and bath could be achieved by  measuring the momentum distribution of the gas or possibly by other methods such as rf-spectroscopy of the SQUID or atomic gas.  This composite system also allows for a myriad of other interesting possibilities. For example, this system could be used to probe the thermalization of quantum systems \cite{EisertNature15}. The flux qubit can be initially prepared in an arbitrary state on the Bloch sphere then brought into resonance with the Fermi gas. This would allow the systematic study of the flow of the initial nonequilibrium state toward the correlated Kondo thermal state. 

The author would like to acknowledge partial support for this project from a Georgia Gwinnett College seed grant. 

\bibliography{/Users/kpatton/Bibliographies/Master}

\end{document}